# Hybridization and Slow Coherence Crossover in the Intermediate Valence Compound YbAl$_3$ via Quasiparticle Scattering Spectroscopy

W. K. Park[1,*], S. M. Narasiwodeyar[1], M. Dwyer[1], P. C. Canfield[2], and L. H. Greene[1]

[1]Department of Physics and the Frederick Seitz Material Research Laboratory, University of Illinois at Urbana-Champaign, Urbana, Illinois 61801, USA

[2]Ames Laboratory and Department of Physics and Astronomy, Iowa State University, Ames, Iowa 50011, USA

Abstract

The intermediate valence compound YbAl$_3$ is known to undergo a hybridization process between itinerant and localized electrons. The resulting heavy Fermi liquid remains non-magnetic and non-superconducting. A microscopic understanding of the hybridization process in YbAl$_3$ is still lacking although some characteristic temperature and energy scales have been identified. Here we report results from novel spectroscopic measurements based on quasiparticle scattering. From the conductance spectra taken over a wide temperature range, we deduce that the band renormalization and hybridization process begins around 110 K, causing the conductance enhancement with a Fano background. This temperature, a new scale found in this work, is much higher than the coherence temperature (34 K). Our observation is in agreement with the slow crossover scenario discussed recently in the literature. The indirect hybridization gap appears to open concomitantly with the emergence of a coherent Fermi liquid. Thus, we suggest its measurement as a more rigorous way to define the coherence temperature than just taking the temperature for a resistivity peak.

*Corresponding author: wkpark@illinois.edu

Most known heavy fermion compounds contain rare-earth or actinide elements as a source for localized moments originating from their 4$f$ or 5$f$ electrons [1]. A canonical model for their underlying physics is the periodic Anderson model, in which the localized moments sit on a lattice and the Hamiltonian describes conduction electrons, localized electrons, and their interaction [1]. A simple one-band hybridization picture, shown in Fig. 1a, captures the essence of band renormalization and hybridization in an Anderson lattice. The localized $f$-band first becomes renormalized around the chemical potential and then hybridizes with the conduction band. This results in two hybridized bands at low temperature, from which a coherent heavy electron liquid emerges, and there appear two kinds of hybridization gap, direct and indirect. These hybridization gaps have been detected via bulk [2-4] and near-surface [5-8] spectroscopies. Recently, we showed that a hybridization gap can also be measured by quasiparticle scattering (or point-contact) spectroscopy (QPS) [9,10]. The indirect gap in the density of states (DOS) can be filled in due to strong correlation or disorder effect [11,12], causing it to look like a pseudogap. However, if the gap does exist, at least a faint trace of its double-peak signature should be detected [10].

Microscopic understanding of the band renormalization and hybridization processes in Anderson lattices is a subject of continued interest [13-16]. At a temperature high enough compared to characteristic temperatures, localized moments are completely decoupled from conduction electrons. As the temperature is lowered, Kondo coupling causes noticeable changes in the charge transport and magnetic properties. For instance, the resistivity in CeCoIn$_5$, a prototype heavy fermion, shows a minimum around 180 K, reminiscent of the single impurity Kondo scattering. With further lowering the temperature, it shows a peak around 45 K and then decreases rapidly [17]. This peak temperature has been referred to as the coherence temperature, $T_{coh}$ (or $T^*$), at which a coherent heavy electron liquid begins to emerge. Not all heavy fermion compounds exhibit such resistivity peaks. In YbAl$_3$, the resistivity decreases monotonically [18,19], whereas its $dc$ magnetic susceptibility shows a peak around $T_{\chi,max}$ ~ 120 K [20,21]. There is an agreement that $T_{coh}$ in YbAl$_3$ is around 30 – 40 K [20], when Fermi liquid behavior appears to set in, but the detailed microscopic processes leading to the coherent state remain to be elucidated. It is also desirable to establish how to define $T_{coh}$ more rigorously in both theory and experiment.

How transport, magnetic, and thermodynamic properties in an Anderson lattice evolve with the tuning of system parameters has been a frequent topic for theoretical investigation [15,16,22,23]. Early on, Schweitzer and Czycholl [22] studied the dependence on the total number of electrons per site, $n_{tot}$, and found the resistivity vs. temperature shows a peak when $n_{tot} \sim 1$ (Kondo lattice regime), and a monotonic temperature dependence if $n_{tot} \ll 1$ (intermediate valence regime). This is in qualitative agreement with experimental observations: e.g., YbAl$_3$, an intermediate valence compound [21,24], i.e., $n_f$ (and $n_{tot}$) $\ll 1$, where $n_f$ is the $f$-level occupancy. Burdin and Zlatić [15] recently studied the effect of the conduction electron DOS on how fast coherence develops; Fig. 1b. YbAl$_3$ would undergo a slow crossover [20,25] to the coherent state as indicated by the track AL1 because the DOS is maximal around the chemical potential. In this case, two temperature scales are necessary to describe the process: Fermi liquid ($T_{FL}$) and single ion Kondo ($T_K$) temperatures. Systems in the Kondo lattice regime, e.g., CeCoIn$_5$, would experience a fast crossover following the track AL2, thus the observed resistivity peak [17]. These contrasting behaviors are summarized by a $T$ vs. $T_{FL}/T_K$ phase diagram in Fig. 1b. This picture provides a way to classify the different behaviors exhibited by various systems. In this Letter, we report our QPS results on YbAl$_3$ providing spectroscopic information on characteristic energy and temperature scales associated with the incoherent-coherent crossover. We also discuss how the emergence of a coherent state is related to the opening of a hybridization gap.

Single crystals of YbAl$_3$ are grown by the self-flux method [18,26]. Its crystalline structure is AuCu$_3$-type cubic as shown Fig. 2 left inset. The resistance data in Fig. 2 show they are of high quality with the residual resistance ratio of $\sim 60$. A simple fit (Fig. 2 right inset) reveals $T_{FL} \sim 34$ K, the Fermi liquid temperature, consistent with the literature [19,20]. The resistivity changes monotonically [18,19] even after subtraction of the non-magnetic contribution [23]; only its first derivative shows a broad peak around 85 K [19]. QPS is a proven spectroscopic technique [27] that can probe electronic properties in the bulk based on a metallic junction by utilizing its ability to inject quasiparticles ballistically [28]. For this, it is crucial to make a clean (metallic) junction whose dimensions are smaller than mean free paths. To overcome the challenge of the YbAl$_3$ crystal surface rapidly developing oxides, we found polishing and Ar ion etching produce the most reproducible junctions. Junctions are formed at low temperature using a differential micrometer [9]. Multiple junctions on different spots are tested *in situ*. Differential conductance is measured as a function of temperature using a standard lock-in technique.

Typical conductance spectra taken at low temperature are shown in Fig. 3. Note the asymmetric background conductance reminiscent of a Fano resonance [9]. Also visible are humps around −15 mV and broad peaks around +10 mV (Figs. 3a & 3b). The background asymmetry and the characteristic energy scales of the hump-peak structure are highly reproducible among junctions of various resistance values (11.8 − 43.6 Ω) as shown in Fig. 3c, strongly suggesting that they are spectroscopic features intrinsic to YbAl$_3$. Speculating that the hump-peak structure could be sharpened in a cleaner (more ballistic) junction [10], we performed extensive QPS measurements using various methods to prepare the sample surface and the tip, but the data shown are representative of the best we could obtain. The hump-peak feature might be smeared due to an intrinsic origin such as strong correlation [12].

Conductance spectra were taken from 4.4 K to 120 K. Since it is somewhat challenging to keep a mechanically formed point-contact junction stable over such a wide temperature range, data were taken in two separate temperature ranges, below and above 30 K, yielding the complete set of spectra plotted in Fig. 4a. The systematic evolution in the conductance shape is more clearly seen in the overlap plot in Fig. 4b. The background asymmetry exists even at the highest temperature. With decreasing temperature, the conductance is gradually enhanced, evolving into the Fano background. To quantify the conductance enhancement, we take a line connecting data points at ±50 mV as a baseline for each conductance curve and numerically integrate the enclosed area. As shown in Fig. 4c, the enhancement begins around 110 K, grows monotonically until it saturates below ~34 K. The hump-peak structure mentioned earlier appears at ~ 30 − 40 K on top of the Fano background.

Two characteristic temperatures can be identified from the conductance spectra: $T_{coh} \sim 34$ K and a new temperature scale, the incoherence temperature, $T_{incoh} \sim 110$ K. $T_{coh}$ matches $T_{FL}$ as derived from our resistance data and in the literature [20]. Our $T_{incoh}$ is in the range where Okamura *et al*. reported a 60 meV kink (or shoulder) in their optical conductivity measurement [29]. $T_{incoh}$ is also close to $T_{\chi,max}$ [20,21]. As already shown in Fig. 3, the hump-peak distance at low temperatures is ~ 25 mV, consistent with the dip energy seen in the optical conductivity at 8 K [29]. This dip energy can be associated with an indirect hybridization gap since it is a minimum energy required for the interband transition (see Fig. 1a). A similar energy scale was also detected in inelastic neutron scattering (INS) measurements [30,31]. Therefore, we reason the hump-peak structure originates from a hybridization gap. All three measurements mentioned here (optical conductivity, INS, and our QPS) are performed without a momentum resolving power, so the gaps mentioned above must be the indirect gap and their quantitative agreement corroborates this conjecture. Angle resolved photoemission spectroscopy (ARPES) can detect band dispersions, hence, both direct and indirect gaps. Earlier ARPES studies on YbAl$_3$ reported a Kondo resonance around the chemical potential [32-34], whose interpretation has been controversial [35-37]. The formation of surface oxides as encountered in our early QPS tests may obscure the observation of the bulk bands using ARPES [38]. Consistent with this is a recent scanning tunneling spectroscopy (STS) study, where results varied spatially as well as with surface preparation and only single impurity Fano resonance features were observed [39].

To elucidate the significance of $T_{incoh}$ and how it is related to $T_{coh}$, we first discuss some phenomenology relevant to the conductance evolution (also, see Fig. 5). Localized moments in an Anderson lattice won't play any role in the charge transport when $T \gg T_K$ since they are completely decoupled from conduction electrons. The conductance curve should then be symmetric and essentially featureless. Thus, the asymmetric conductance observed even at the highest temperature (120 K) in Fig. 4 indicates that this temperature is still below $T_K$, in agreement with the literature ($T_K \sim 670$ K) [20]. The reason why our QPS conductance shows such an asymmetry instead of the typically observed symmetric curve with a dip around zero bias as seen in QPS measurements on single impurity Kondo alloys [27] may be due to a Fano interference effect [9]. After all, localized moments in an Anderson lattice are not scarce near the junction area, so it is natural for quasiparticles to interfere as they are injected via two scattering channels, one into the conduction band and the other into the localized states, the so-called Kondo-Fano resonance [9]. Note that similar high-temperature asymmetries are also observed in our QPS studies on other heavy fermions [28,40,41]. While $T_K$ indicates the formation of local Kondo singlets, we argue that $T_{incoh}$ signifies the beginning of their coupling throughout the lattice. This is consistent with our earlier discussion (and Fig. 1b): in the Kondo lattice regime, the intersite coupling occurs collectively leading quickly to a coherent state; but in YbAl$_3$, which is in the intermediate valence regime, the coherence occurs slowly [20,25] as detailed below. The conductance enhancement grows continuously below $T_{incoh}$, forming the Fano background with a broad peak centered at a positive bias until the additional hump-peak structure develops around $T_{coh}$.

In Fig. 5, we illustrate a scenario for the hybridization process in YbAl$_3$. The indirect gap plotted is based on the estimated peak-hump distances. The schematic plot for the direct gap is based on the features observed in optical conductivity [29], that is, the 60 meV kink (or shoulder) seen below 80 − 120 K. These hybridization gaps could be extracted from a more formal analysis [11,12,42,43] as done in our earlier work on URu$_2$Si$_2$ [9] but we opted not to do that here since i) the hump-peak structure is not much pronounced and ii) such an analysis is not crucial to our main reasoning. The insets in Fig. 5 show speculated band diagrams and simulated conductance curves at different stages in the hybridization process. Notice that the two characteristic temperature scales are marked: $T_{coh}$ ~ 34 K and $T_{incoh}$ ~ 110 K. Here we can see $T_{coh} = T_{FL}$, consistent with the coherent Fermi liquid being formed at this temperature [20]. We further argue that $T_{coh} = T_{hyb}$ since the indirect hybridization gap appears to open around $T_{coh}$ as noted earlier. The other temperature scale, $T_{incoh}$, defines when an incoherent coupling between the renormalized *f*-band and the conduction band begins. Microscopically, this means that the hybridization strength (*V*), thus, also the direct gap (2*V*), becomes finite at $T_{incoh}$, as indicated in Fig. 5. This new temperature scale is distinct from $T_K \sim 670$ K [20]. For $T_{incoh} < T < T_K$, there is no intersite coupling among Kondo singlets but only local Kondo screening. The intersite coupling that begins at $T_{incoh}$ doesn't lead to coherent hybridized bands immediately. Instead, this happens at a much lower temperature, $T_{coh} = T_{hyb,} = T_{FL}$, when a hybridization gap opens and a coherent heavy Fermi liquid emerges.

Our results along with other reports [20,25] establish that the incoherence-coherence crossover in YbAl$_3$ occurs slowly. Burdin and Zlatić [15] showed that this slow crossover results from the conduction electron DOS being maximal at the chemical potential. We argue that this picture could be elaborated further by defining the coherence temperature more rigorously than just taking the resistivity peak temperature. As we noted already [9], a peak in resistivity indicates that one kind of scattering dominates over the other crossing the peak temperature. Without detailed information of existing scattering processes, taking the resistivity peak temperature is not a well-defined way to determine $T_{coh}$. As shown here, finding the temperature at which a hybridization gap opens would be a more rigorous way. In accord with this, we speculate that $T_{coh}$ in URu$_2$Si$_2$ is higher than 34 K since gap-like signatures are still observed at this temperature in our QPS [10]. It is interesting to note that, theoretically, the hybridization is complete or the full coherence is established when all the slave bosons are condensed [44] and the gap opens. In practice, determining the hybridization gap could be a challenging task due to complicated involvement of crystal field splitting, spin-orbit coupling, multi-orbital nature, etc.

In conclusion, QPS on YbAl$_3$ crystals leads to a microscopic understanding of how the coherence develops in this intermediate valence compound. The conductance enhancement with a Fano background beginning around 110 K signifies the incoherent coupling among the Kondo singlets, and a full coherence in the hybridized bands, or a hybridization gap, is not achieved until 34 K, supporting the slow crossover picture in the literature. The hybridization gap we find of ~25 meV is also consistent with that observed by other spectroscopic measurements. We suggest a new way to determine the coherence temperature more rigorously, namely, by finding the temperature at which a hybridization gap opens.

We thank P. Saraf for the help with QPS, J. Frederick for the crystal growth, and R. Haasch for the help with surface characterization and analysis. This material is based upon work supported by the U.S. NSF DMR under Award 12-06766; work done at Ames Laboratory is supported under DOE Contract No. DE-AC02-07CH11358.

**FIGURE CAPTIONS**

**Figure 1. a**, Simple one-band hybridization picture for the periodic Anderson model. The conduction band ($\varepsilon_k$) and the renormalized *f*-band ($\lambda$) hybridize, resulting in two hybridized bands. $\mu$ denotes the chemical potential. Also indicated are two hybridization gaps, direct ($\Delta_{dir} = 2V$) and indirect ($\Delta_{hyb} = 2V^2/D$; $2D$ = conduction band width). **b**, Schematic phase diagram accounting for the incoherent-coherent crossover process in an Anderson lattice, adapted from Ref. [15]. $T_K$ is the single ion Kondo temperature and $T_{FL}$ is the temperature scale for a Fermi liquid. Tracks labeled AL1 and AL2 contrast systems undergoing a fast and a slow crossover, respectively.

**Figure 2.** Resistance data indicate the high quality of the YbAl$_3$ crystals (residual resistance ratio of ~60). Upper inset: AuCu$_3$-type cubic crystal structure for YbAl$_3$. Lower inset: Resistance vs. $T^2$ plot, giving $T_{FL}$ ~ 34 K.

**Figure 3. a & b**, Typical QPS conductance data at low temperature (4.3 − 4.6 K). The most noticeable feature is a Fano-like background asymmetry. Humps around −15 mV and peaks around +10 mV as indicated by arrows are reproducibly observed. A slight signature for an additional hump around zero-bias (the asterisk) is sometimes visible. **c**, Collection of conductance curves for junctions with the resistance at −50 mV ranging from 11.8 − 43.6 Ω. Not only is the background asymmetry reproducible but also the hump-peak structure is visible in all junctions around similar bias voltages as indicated by the dotted lines, suggesting that they are spectroscopic features intrinsic to YbAl$_3$.

**Figure 4. a**, Conductance spectra taken from 4.36 K to 119.64 K, normalized at −50 mV, and shifted vertically for clarity. **b**, The same spectra plotted without shifting. **c**, Excess conductance as determined by the area under each conductance curve above a linear baseline connecting the data points at ±50 mV.

The conductance enhancement begins around 110 K and saturates below ~34 K. The fluctuating behavior below ~30 K is due to re-forming the junctions in that region.

**Figure 5.** Schematic plots to illustrate how the band renormalization and hybridization process occurs in YbAl$_3$. The hybridization process occurs so slowly that a coherent Fermi liquid emerges only at $T_{coh}$ = 34 K (<< $T_{incoh}$ = 110 K). The indirect gap schematically plotted is based on the estimated hump-peak distances at low temperatures. The schematic plot for the direct gap is for the qualitative discussion in the text. As indicated, we argue that $T_{FL} = T_{coh} = T_{hyb}$, where $T_{hyb}$ is the indirect gap opening temperature. The insets are speculated band diagrams and simulated conductance curves at different stages of the process.

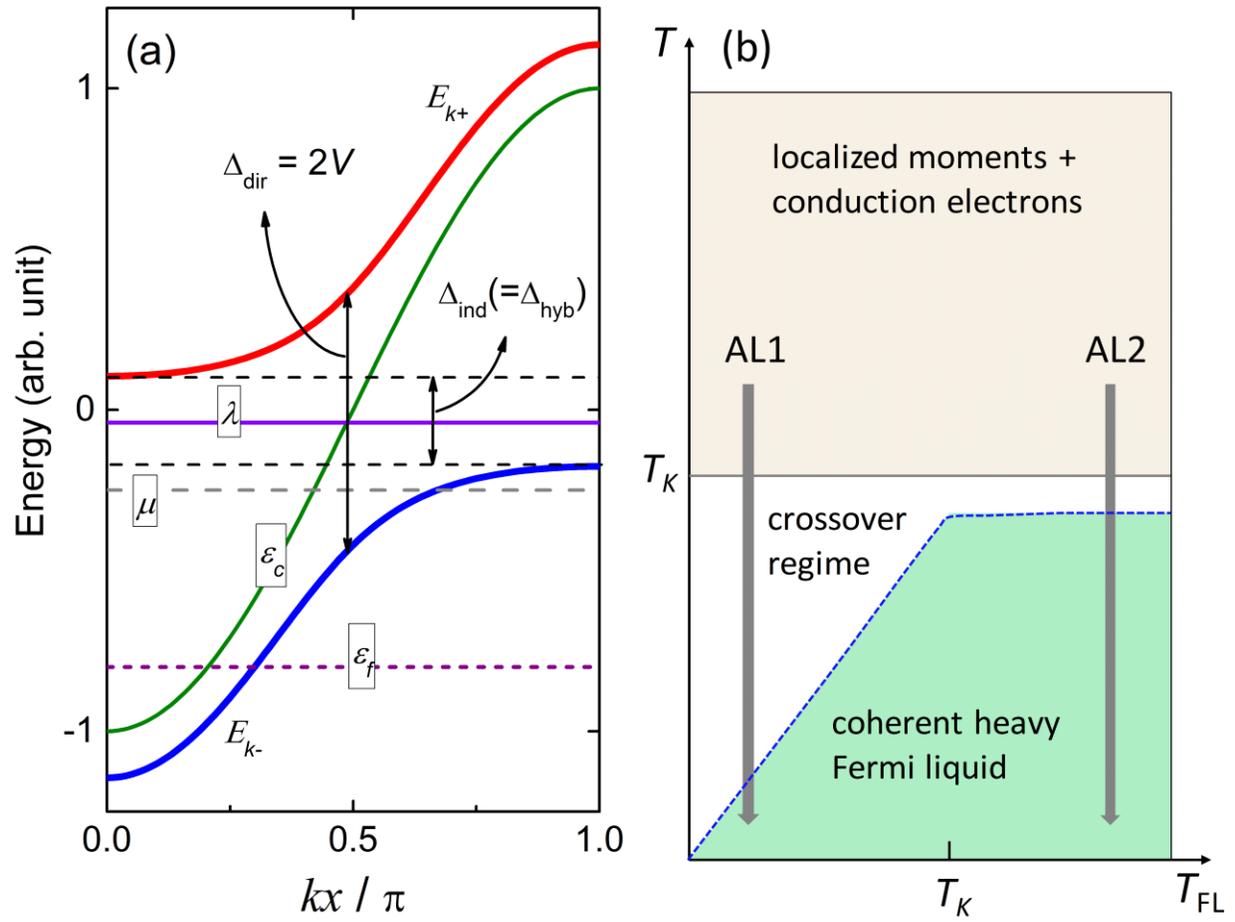

Figure 1. W. K. Park *et al*.

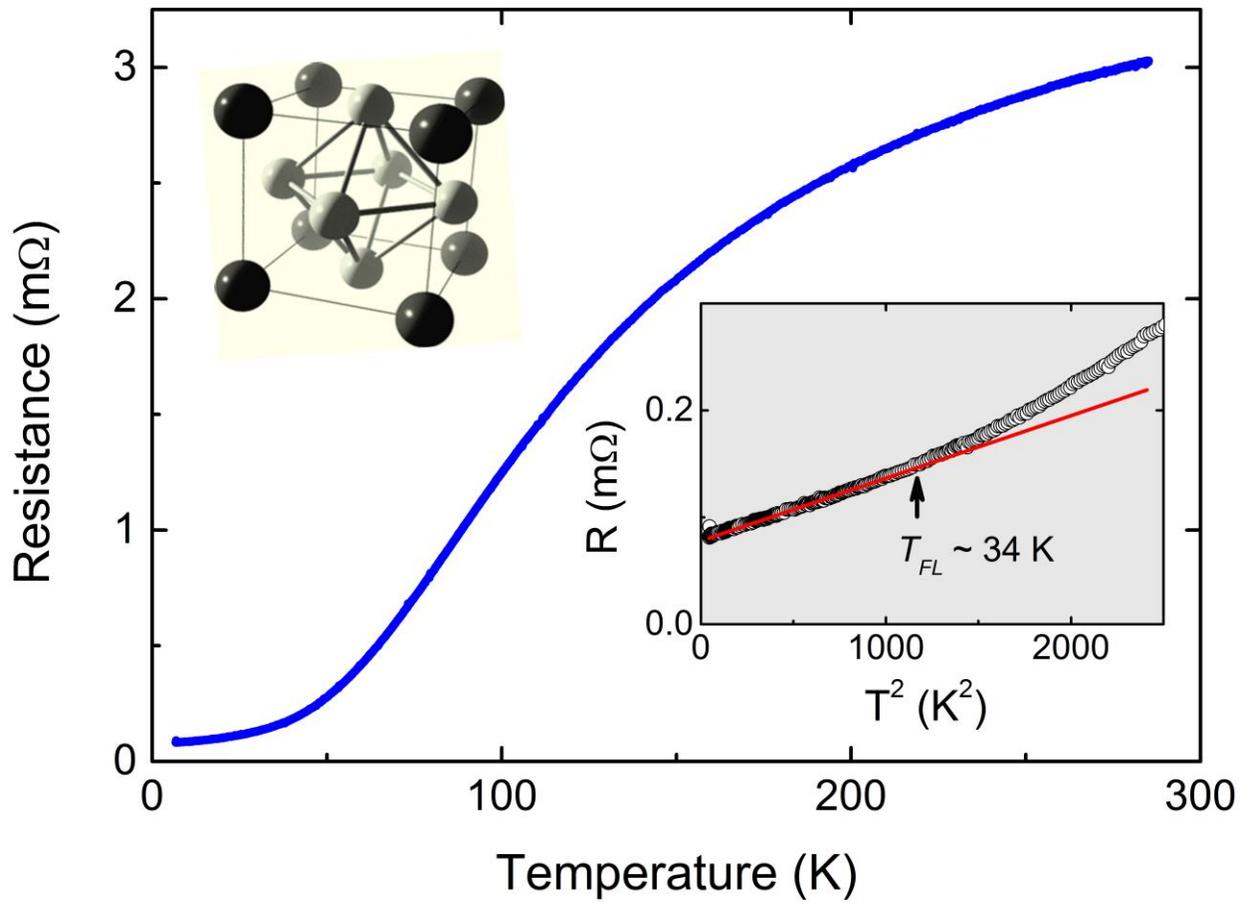

Figure 2. W. K. Park *et al*.

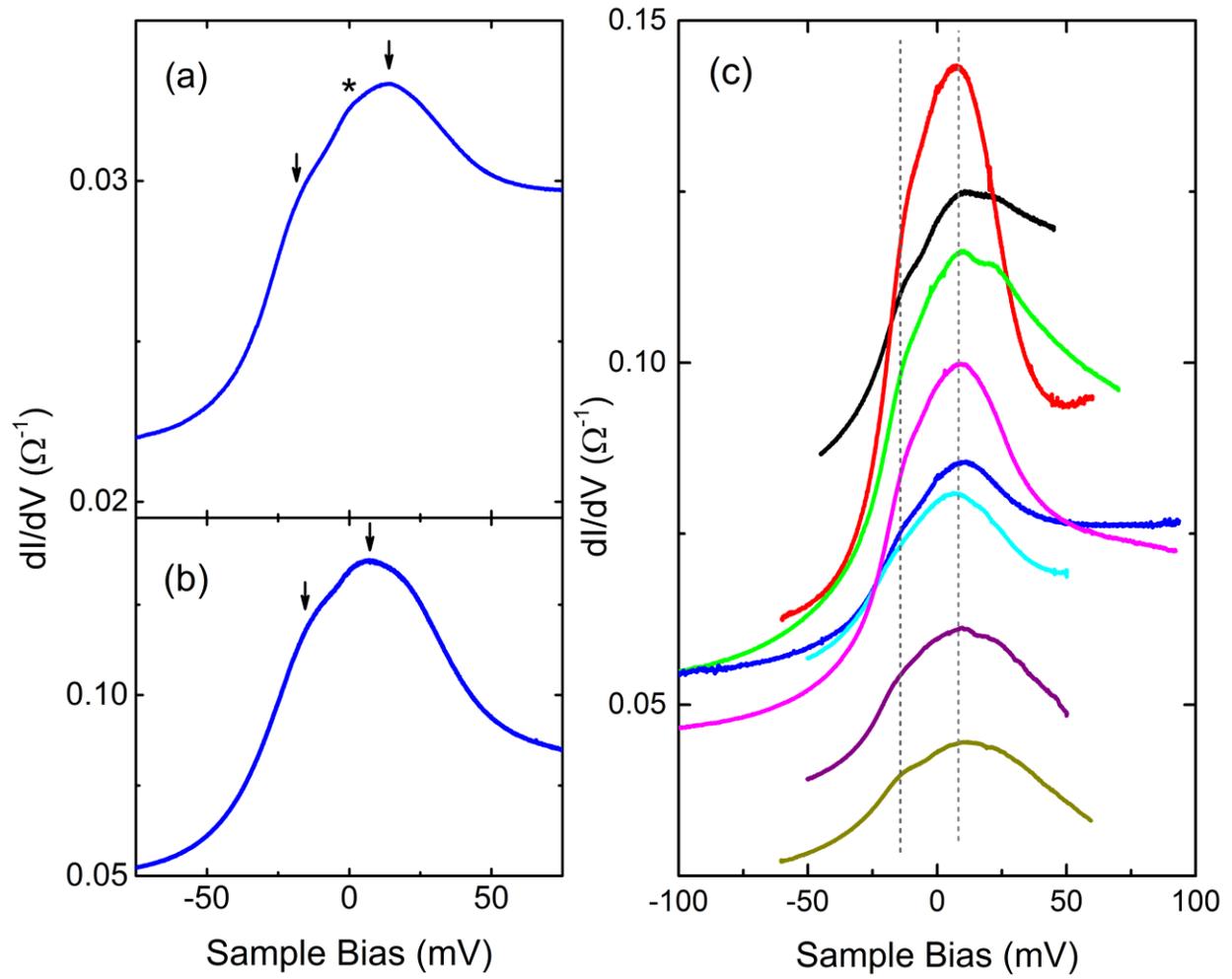

Figure 3. W. K. Park *et al.*

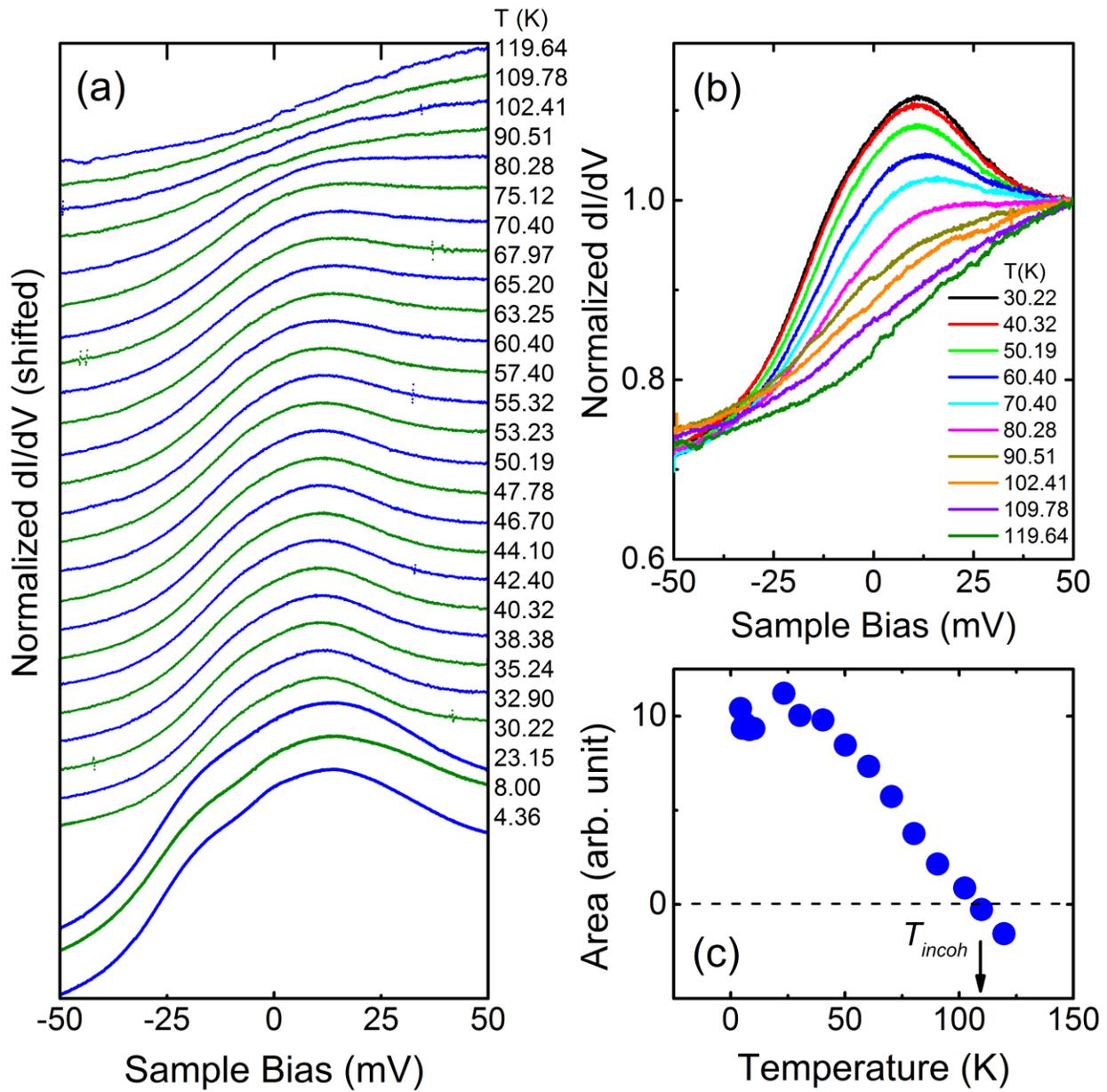

Figure 4. W. K. Park *et al*.

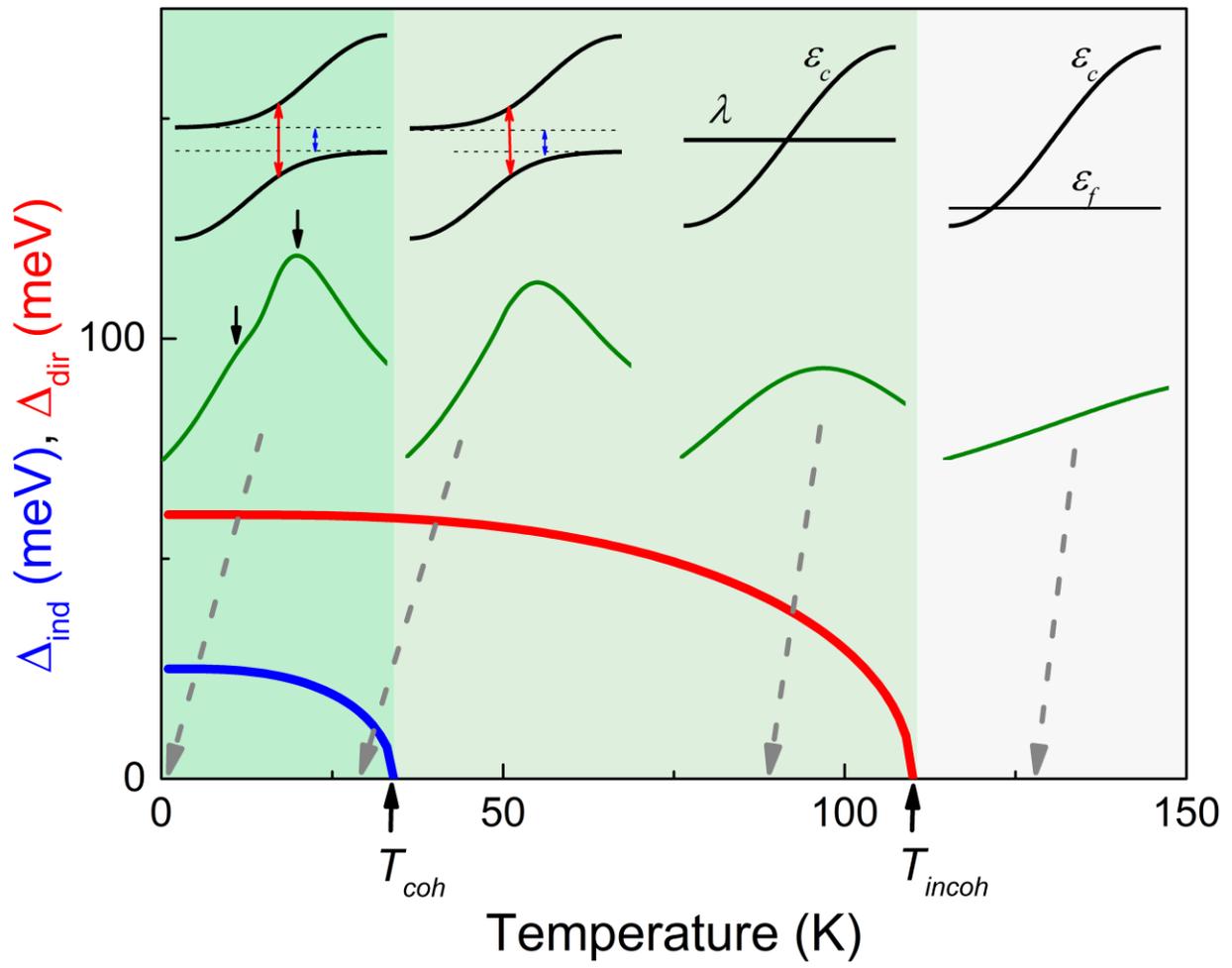

Figure 5. W. K. Park *et al.*